 \documentclass[final,5p,times,twocolumn]{elsarticle}

\usepackage{amssymb}
\usepackage{lipsum}
\usepackage[hyphens]{url}  
\usepackage{breakurl}
\usepackage{hyperref}
\usepackage{etoolbox} 
\usepackage{balance}
\usepackage{amsmath}
\usepackage{array}
\usepackage{booktabs}
\usepackage{tikz}
\usepackage{changepage}
\usetikzlibrary{shapes.geometric, arrows.meta, positioning}

\journal{Pre-print}

\newcommand{\orcidlink}[1]{\textsuperscript{\href{https://orcid.org/#1}{\includegraphics[scale=0.2]{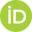}}}}

\begin{document}

\begin{frontmatter}



\title{Virtual Reality in Sign Language Education: Opportunities, Challenges, \\and the Road Ahead}


\author[]{Santiago Berrezueta-Guzman\,\orcidlink{0000-0001-5559-2056}}
\author[]{Refia Daya\,\orcidlink{0009-0001-9546-3950}}
\author[]{Stefan Wagner\,\orcidlink{0000-0002-5256-8429}}
\affiliation[]{organization={Technical University of Munich},
           city={Heilbronn},
            country={Germany}}

\begin{abstract}
\begin{center}
\begin{minipage}{0.9\textwidth}
Sign language (SL) is an essential mode of communication for Deaf and Hard-of-Hearing (DHH) individuals. Its education remains limited by the lack of qualified instructors, insufficient early exposure, and the inadequacy of traditional teaching methods. Recent advances in Virtual Reality (VR) and Artificial Intelligence (AI) offer promising new approaches to enhance sign language learning through immersive, interactive, and feedback-rich environments. This paper presents a systematic review of 55 peer-reviewed studies on VR-based sign language education, identifying and analyzing five core thematic areas: (1) gesture recognition and real-time feedback mechanisms; (2) interactive VR environments for communicative practice; (3) gamification for immersive and motivating learning experiences; (4) personalized and adaptive learning systems; and (5) accessibility and inclusivity for diverse DHH learners.

The results reveal that AI-driven gesture recognition systems integrated with VR can provide real-time feedback, significantly improving learner engagement and performance. However, the analysis highlights critical challenges: hardware limitations, inconsistent accuracy in gesture recognition, and a lack of inclusive and adaptive design. This review contributes a comprehensive synthesis of technological and pedagogical innovations in the field, outlining current limitations and proposing actionable recommendations for developers and researchers. By bridging technical advancement with inclusive pedagogy, this review lays the foundation for next-generation VR systems that are equitable, effective, and accessible for sign language learners worldwide.
\end{minipage}
\end{center}
\end{abstract}



\begin{keyword}
\begin{center}
\begin{minipage}{0.9\textwidth}
Virtual Reality \sep Sign Language Education\sep Gesture Recognition\sep Deaf and Hard-of-Hearing (DHH)\sep AI in Learning\sep Immersive Learning Environments\sep Accessibility and Inclusivity.


\end{minipage}
\end{center}
\end{keyword}

\end{frontmatter}

\section{Introduction}

Sign languages (SLs) are natural human languages with their own phonology, morphology, syntax, and pragmatics, born and evolving within Deaf communities around the world. Although they make use of the visual–spatial modality rather than the auditory–vocal channel, sign languages are linguistic systems just as rich and expressive as spoken languages, and in no way are they simple gestural codes or mere translations of the surrounding spoken tongue \citep{stokoe2005sign, klima1979signs}. Each SL, whether American Sign Language (ASL), British Sign Language (BSL), or Brazilian Sign Language (LIBRAS), to name a few, developed independently, with unique grammatical structures and regional variation, entirely independent of the ambient spoken language.  

As with any natural language, fluency in a sign language grants its users full access to education, employment, social services, and cultural life. Yet, despite official recognition in many countries, SL education remains unevenly available: globally, only about 5–10\% of Deaf and Hard-of-Hearing (DHH) children acquire a sign language from birth, while the vast majority encounter their first structured exposure to SL much later, often through under-resourced school programs or self-directed learning \citep{G-NEWPORT1988147, humphries2012language}. These delays and limitations in early access have profound effects on cognitive and linguistic development, on educational attainment, and on social inclusion. Addressing these inequities requires both expanding access to qualified instructors and exploring innovative pedagogical approaches—including the use of immersive and interactive technologies such as Virtual Reality (VR) and Artificial Intelligence (AI)—to bring high-quality, accessible SL learning to all who need it.

A significant portion of the DHH population learns SL later in life due to a lack of early exposure, the scarcity of qualified instructors, and the limited integration of SLs into formal education systems \cite{D-9072479, K-article}. Traditional teaching approaches, such as textbooks, video lessons, and instructor-led classrooms, often fail to capture SL's visual-spatial and expressive complexities, limiting learners' engagement, comprehension, and retention \cite{H-wen2024enhancingsignlanguageteaching, J-article, O-unknown}.

Recent advancements in immersive technologies, particularly the integration of Virtual Reality (VR) and Artificial Intelligence (AI), offer new opportunities for transforming SL education \cite{F-10536554, ozkaya2025llms}. VR enables three-dimensional, interactive, and context-rich environments that simulate real-life communication scenarios, provide visual demonstrations from multiple perspectives, and support real-time feedback \cite{A-alam2023aslchampvirtualreality, U-article, C-inproceedings}. When enhanced with AI-powered gesture recognition, these systems can analyze learner performance, deliver personalized feedback, and create gamified, adaptive experiences that support diverse learners \cite{B-10.1145/3411763.3451523, 4.1.o-article}. Despite this promise, current implementations remain fragmented, with notable limitations in scalability, accessibility, cultural inclusivity, and pedagogical depth \cite{D-9072479, 4.1.l-article, 4.1.q-sabbella2024evaluatinggesturerecognitionvirtual}.

This paper presents a systematic review of VR-based SL education. It synthesizes findings across 55 peer-reviewed studies to identify the key applications, technological innovations, and pedagogical strategies employed in the field. This review focuses on five thematic areas: (1) gesture recognition and feedback mechanisms; (2) interactive VR environments for practice; (3) immersive learning through gamification; (4) personalized learning and adaptive systems; and (5) accessibility and inclusivity.

By mapping these areas against the current challenges and future needs of learners and educators, this paper informs the design of next-generation learning tools that are effective, ethical, and equitable. Additionally, it provides practical recommendations for developers, educators, and researchers and proposes directions for future work to support scalable, inclusive, and culturally responsive SL learning through immersive technologies.

\section{Background}

The intersection of SL education and emerging immersive technologies has gained increasing attention as educators and researchers seek innovative ways to address persistent challenges in accessibility, pedagogy, and learner engagement.

\subsection{Overview of Sign Language Education}

Sign languages are acquired through distinct developmental trajectories depending on learners' age, hearing status, and learning context. Deaf children born to signing parents (approx.\ 5–10 \% of the population) acquire a sign language natively, following the same milestones as hearing children do in spoken language acquisition \citep{schick2010development}. Early exposure—whether from Deaf family members, early intervention programs, or signing daycare environments—ensures robust linguistic and cognitive development, and has become increasingly available through universal newborn hearing screening and family‐centred early intervention services \citep{lillo2021acquisition}.

For deaf children of hearing parents, structured SL instruction often begins in infancy or toddlerhood via early intervention specialists and specialized preschool programs, rather than waiting for the first school classroom. Contemporary models integrate Deaf mentors, parent coaching, and bilingual–bicultural curricula, recognizing that home signing systems (‘‘home signs’’) lack the full grammatical complexity of natural SLs but can scaffold early communication until richer input is provided \citep{branson2005damned, morford2011homesigners}.  

By contrast, hearing adult learners—including parents, interpreters, and professionals—typically begin structured SL study well after childhood. Adult courses range from university credit classes and community workshops to online modules and immersion camps. Pedagogical approaches for adults emphasize explicit instruction in SL grammar (handshape, location, movement, non-manual markers), cultural norms, and functional fluency, often combining video-based modeling, in-person practice with Deaf signers, and increasingly, digital platforms with interactive feedback \citep{quinto2011teaching}. Research shows that adult learners benefit from multimodal input and scaffolded practice, but progress varies widely based on intensity, motivation, and access to native models.

Across all learner groups, effective SL education requires  
\begin{itemize}
  \item \emph{Rich linguistic input} from fluent signers (for both child and adult learners);  
  \item \emph{Immediate, targeted feedback} on phonological parameters and facial grammar;  
  \item \emph{Culturally responsive pedagogy} that situates signs within Deaf community norms and Deaf–hearing intercultural communication.  
\end{itemize}
Despite expanded early intervention and adult instruction options, gaps remain in qualified Deaf instructor availability, equitable program funding, and evidence‐based curriculum standards—challenges that immersive technologies like VR/AI are uniquely poised to address.  

For the deaf community, SL is a fundamental means of communication. Despite its importance, learning and teaching SL remain challenging due to limited course availability, a shortage of qualified instructors, and the language's inherent complexity. For example, American Sign Language (ASL) is a visual-gestural language with unique linguistic structures. However, unlike most spoken languages, only about 5-10\% of deaf individuals acquire ASL from birth in households where SL is naturally used. The majority are instead introduced to it later in life, often through formal education programs rather than through natural exposure in their home environments \cite{G-NEWPORT1988147}.

Although SLs are recognized as official languages in many countries, the levels of fluency are low because of the barriers to acquisition and the absence of practical approaches to teaching \cite{D-9072479}. Most students learn SL for the first time at school and not at home, and, historically, SL was banned from the classroom, with deaf children having to teach themselves from other deaf children \cite{K-article}. 

SL education often relies on classroom teaching, textbooks, and video-based resources. While classroom instruction can provide personalized feedback, it is frequently limited by a lack of qualified instructors and difficulty offering consistent one-on-one support \cite{O-unknown}. Textbooks and videos are more flexible and accessible but lack interactivity and fail to represent the spatial and expressive features of SL, such as hand orientation, movement, and facial expressions, that are essential for accurate communication \cite{H-wen2024enhancingsignlanguageteaching, J-article}. 
Research indicates that while some learners are fortunate to be taught by native signers, many are instructed by non-native users, which introduces variability in teaching effectiveness \cite{G-NEWPORT1988147}.

Approximately 90--95~\% of deaf children are born to hearing parents who do not use SL, resulting in limited or no exposure to a fully developed linguistic system during critical developmental years. While some children develop 'home signs'—gestural systems created within families to facilitate essential communication—these systems lack the grammatical complexity of natural languages and do not support complete language acquisition. This early lack of linguistic input can lead to cognitive and language development delays, making subsequent mastery of a formal SL significantly more difficult \cite{G-NEWPORT1988147}.

Moreover, feedback is critical in SL acquisition. Despite its importance, there is currently no standardized approach to integrating feedback into SL learning methodologies \cite{J-article, J1-article, K-article}. Learners are typically required to self-monitor and self-correct, which is particularly difficult in the absence of visual or interactive cues \cite{A-alam2023aslchampvirtualreality, H4-inproceedings, H5-inproceedings, H6-inproceedings}. While some technological solutions aim to address this gap, current SL recognition systems still face challenges in accurately interpreting dynamic gestures and facial expressions—both essential components of meaning in SL communication \cite{H-wen2024enhancingsignlanguageteaching}.

\subsection{Virtual Reality in Sign Language Education}

Initially developed for entertainment, Virtual Reality (VR) has increasingly gained traction in educational contexts and is now applied across various domains. Its growing relevance in education stems from its ability to simplify complex concepts and extend learning opportunities beyond geographical boundaries \cite{A-alam2023aslchampvirtualreality}.

In language learning, VR supports the development of communicative competence, cultural awareness, critical thinking, and kinesthetic understanding, particularly beneficial when learning spatially grounded languages such as SL \cite{U-article}. VR also facilitates real-time interaction and task-based learning, allowing learners to practice in realistic scenarios and perform authentic communicative tasks \cite{ chen2025task}.

Furthermore, studies show that VR enhances motivation and learner autonomy, with serious games and gamified experiences contributing to higher engagement and persistence compared to traditional methods such as lectures or video-based instruction \cite{J2-inproceedings, J3-article, damianova2025serious, sobchyshak2025pushing}. 

Aligned with experiential learning theory, VR fosters active learning by immersing students in problem-solving and decision-making tasks, which supports memory retention, deeper comprehension, and the development of cognitive and metacognitive skills \cite{D-9072479, U-article}. This is particularly valuable in SL education, where learners must grasp spatial relationships, precise hand movements, and visual cues. Traditional methods often lack interactivity and fail to represent the dynamic, three-dimensional nature of SL \cite{F-10536554, K-article}.

A core advantage of VR in this context is its integration with motion capture technology, which enables real-time gesture recognition and immediate feedback on accuracy. Devices such as Leap Motion controllers and Oculus hand tracking enhance the ability of VR applications to track hand positions and analyze sign execution \cite{C-inproceedings, D-9072479}. These tools ensure learners receive precise visual feedback, helping them correct errors and internalize accurate signing techniques.

Machine learning models, including Convolutional Neural Networks (CNNs) and Hidden Markov Models (HMMs), are employed to interpret gestures and evaluate performance in real-time. These models allow learners to practice signs in VR environments where errors are detected automatically and corrective feedback is provided instantly. These algorithms assess hand postures and inform the user whether the gesture was performed correctly, encouraging repetition and reinforcement \cite{A-alam2023aslchampvirtualreality, B-10.1145/3411763.3451523}.

To enhance realism and instructional support, some VR systems integrate signing avatars—driven by motion‐capture or advanced pose‐estimation tools such as AlphaPose, Azure Kinect, and MediaPipe—to demonstrate correct sign execution and prompt real-time user adjustments \citep{A-alam2023aslchampvirtualreality}. For example, Bansal et al.’s CopyCat uses these vision-only pipelines alongside Hidden Markov Models trained on keypoint data to verify ASL sentence production in real time, achieving over 90 \% word‐level accuracy without specialized gloves \citep{B-10.1145/3411763.3451523}. Such AI-powered avatar feedback loops substantially improve comprehension, retention, and practical SL application in immersive learning environments.

Some applications often simulate everyday conversations and social scenarios, bridging the gap between theoretical instruction and real-life communication \cite{K-article}. Such contextual learning not only reinforces vocabulary but also supports the development of fluency and situational understanding \cite{O-unknown}.

Engagement in VR-based SL education is further enhanced through gamification and adaptive learning strategies. Studies have shown that game-like environments increase learner motivation, persistence, and time spent in learning compared to traditional formats \cite{F-10536554, O2-article, O4-10.1145/3640544.3645218}. AI-driven avatars within these environments often serve as interactive tutors, offering immediate performance feedback and visual scoring systems to support learner understanding \cite{C-inproceedings, O6-10.1007/978-3-319-09635-3_12}.

Although VR for SL education is still evolving, ongoing developments suggest a promising future that includes haptic feedback to reinforce correct handshapes and multi-modal recognition systems capable of interpreting facial expressions—an essential component of SL grammar and meaning \cite{A-alam2023aslchampvirtualreality, H-wen2024enhancingsignlanguageteaching}.

\section{Methodology}

This study employed a systematic literature review to examine current applications of Virtual Reality (VR) in Sign Language (SL) education. The methodology involved a structured search strategy, selection criteria, and categorization to identify key themes and trends across recent works. 

\subsection{Literature Search Strategy}

Our search included electronic databases such as IEEE Xplore, ACM Digital Library, Scopus, Web of Science, and Google Scholar. The search terms used were combinations of keywords such as "\textit{sign language learning}," "\textit{virtual reality}," "\textit{VR-based education,}" "\textit{gesture recognition}," "\textit{deaf education}," and "\textit{interactive environments}." Boolean operators (\texttt{AND}, \texttt{OR}) were applied to refine the search.

The inclusion criteria were:
\begin{itemize}
    \item Studies published between 2010 and 2025.
    \item Peer-reviewed journal articles, conference proceedings, or book chapters.
    \item Studies that include technical or pedagogical evaluations of VR systems in SL learning.
\end{itemize}

Exclusion criteria included:
\begin{itemize}
    \item Studies not written in English.
    \item Works that did not involve VR or SL learning directly.
    \item Purely theoretical or opinion papers without implementation or user evaluation.
\end{itemize}

\begin{figure}[ht]
\centering
\begin{tikzpicture}[
    node distance=0.3cm and 0.6cm,
    every node/.style={font=\small, align=center},
    box/.style={rectangle, draw=black, thick, text width=3.5cm, minimum height=1cm, align=center},
    arrow/.style={-{Latex[length=2mm]}, thick}
]

\node[box] (id) {Records identified through database searching\\(n = 347)};
\node[box, below=of id] (dup) {Records after duplicates removed\\(n = 255)};
\node[box, below=of dup] (screen) {Records screened (title/abstract)\\(n = 255)};
\node[box, right=of screen, text width=4.5cm] (excl1) {Records excluded\\(n = 141)};
\node[box, below=of screen] (fulltext) {Full-text articles assessed for eligibility\\(n = 114)};
\node[box, right=of fulltext, text width=4.5cm] (excl2) {Full-text articles excluded (e.g., no user evaluation, not SL-focused)\\(n = 59)};
\node[box, below=of fulltext] (incl) {Studies included in review\\(n = 55)};

\draw[arrow] (id) -- (dup);
\draw[arrow] (dup) -- (screen);
\draw[arrow] (screen) -- (fulltext);
\draw[arrow] (fulltext) -- (incl);
\draw[arrow] (screen.east) -- (excl1.west);
\draw[arrow] (fulltext.east) -- (excl2.west);

\end{tikzpicture}
\caption{PRISMA flow diagram illustrating the study selection process.}
\label{fig:prisma}
\end{figure}
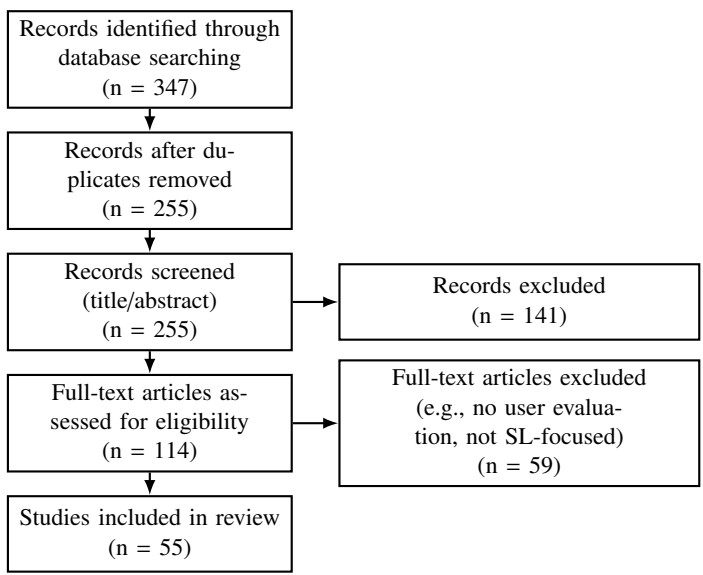

\begin{figure*}[h]
    \centering
    \includegraphics[width=1\textwidth]{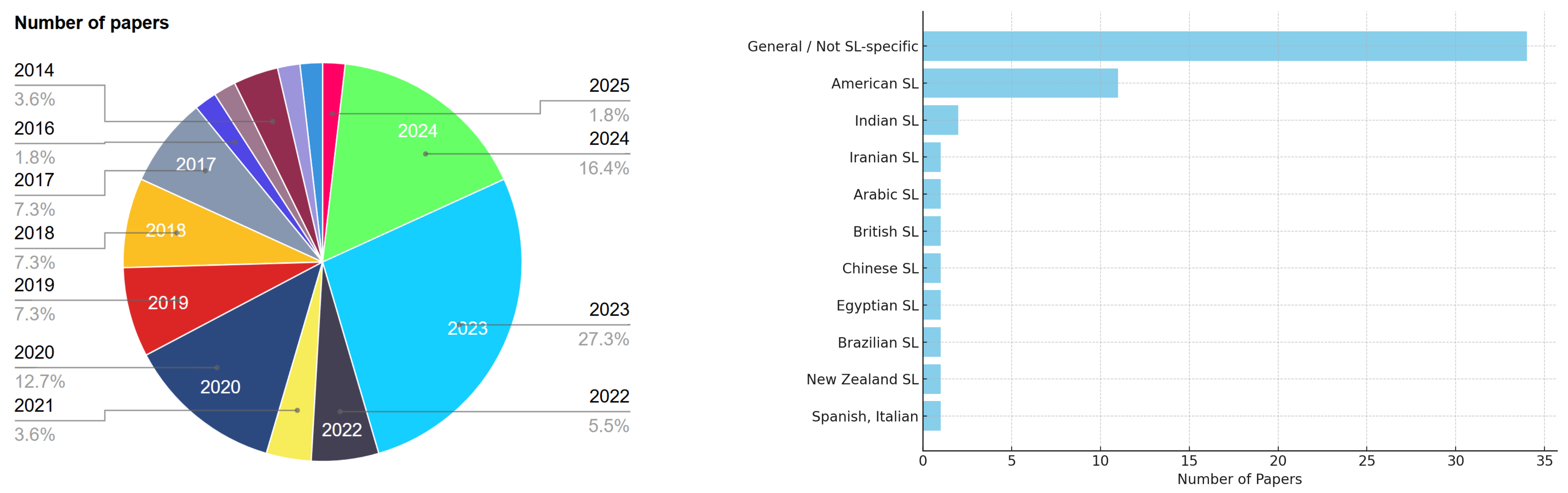} 
    \caption{Distribution of the analysed papers by their publication years and sign language, respectively.}
    \label{fig1}
\end{figure*}

Figure~\ref{fig:prisma} provides a PRISMA flowchart which shows that 55 relevant studies were identified for full-text analysis, and Figure~\ref {fig1} illustrates how these studies are distributed by year and type of language. 

\subsection{Categorization of Literature}

Each article was reviewed for its objectives, methodology, technology, and findings. Five major categories emerged from this analysis:

\begin{enumerate}
    \item \textit{Gesture Recognition and Feedback Mechanisms:} Studies focusing on AI/ML-based recognition systems for sign detection and real-time feedback.
    \item \textit{Interactive VR Environments for Practice:} Applications that simulate conversational or task-based scenarios for SL use.
    \item \textit{Immersive Learning Through Gamification:} Studies implementing game elements to increase engagement and retention.
    \item \textit{Personalized Learning and Adaptive Systems:} Systems offering user-specific content progression, replay features, or real-time adaptation.
    \item \textit{Accessibility and Inclusivity:} 
    Works addressing the needs of DHH users through avatars and translation tools.
\end{enumerate}

\section{Key Applications of Virtual Reality in Sign Language Learning}

The final categorization enabled a structured synthesis of the findings and facilitated the construction of Table~\ref{tab:vr-sign-language-summary} that summarizes principal applications, limitations, and key studies.

\begin{table*}[ht]
\centering
\caption{Summary of Key Findings and Limitations of Virtual Reality in Sign Language Learning.}
\begin{tabular}{p{2.3cm} p{7.3cm} p{7cm}}
\toprule
\textbf{Category} & \textbf{Principal Findings} & \textbf{Limitations} \\
\midrule
\textbf{Gesture recognition and feedback mechanisms}  & 
Advanced gesture recognition using CNNs, LSTMs, HMMs, and 3D CNNs enables real-time sign evaluation with high accuracy; visual feedback facilitates learning; Unity and Oculus are commonly used. & 
Hardware dependency, limited gesture datasets, no haptic feedback, challenges with dynamic gestures, and low facial expression tracking accuracy. \\
\midrule
\textbf{Interactive VR environments for practice }& 
Gamified VR with contextual scenarios enables immersive practice; avatar-based multiplayer design fosters realistic interactions and engagement. & 
It mostly uses individual use, has limited multiplayer integration, and lacks culturally adapted communication models. \\
\midrule
\textbf{Immersive learning through gamification } & 
Gamification improves motivation, engagement, and retention; gesture-triggered puzzles and animations reinforce sign-memory links. & 
Some gamified elements are superficial; they may not address learner autonomy, confusion in 3D environments, or match novice users' needs. \\
\midrule
\textbf{Personalized learning and adaptive systems}  & 
Some systems allow manual replay, camera control, and non-linear navigation; conceptual frameworks for adaptive learning exist. & 
Lack of AI-based adaptation, minimal real-time performance tracking, and no intelligent feedback loops. \\
\midrule
\textbf{Accessibility and inclusivity}  & 
VR avatars and visual feedback support DHH learners; they report improved motivation, and the proposed multilingual environments. & 
High hardware cost, limited support for interpreter avatars, absence of standardized academic signs, and limited support for users with multiple disabilities. \\
\bottomrule
\end{tabular}
\label{tab:vr-sign-language-summary}
\end{table*}

\subsection{Gesture Recognition and Feedback Mechanisms}

AI-driven gesture recognition in VR-based SL learning focuses on accurately detecting hand shapes, movements, and spatial configurations. Convolutional Neural Networks (CNNs) are commonly used for static gesture recognition, while Long Short-Term Memory (LSTM) networks are more effective for modeling temporal dynamics \cite{4.1.a-unknown, 4.1.c-unknown, 4.1.m-8623035, 4.1.n-article}. Hidden Markov Models (HMMs) have also shown strong performance, particularly in structured, real-time educational settings involving young learners, with one system achieving over 90~\% user-independent accuracy and outperforming transformer-based models \cite{A-alam2023aslchampvirtualreality, B-10.1145/3411763.3451523}. More advanced models, including 3D CNNs, deformable 3D CNNs, and convolutional LSTMs, have been applied to classify gestures in video and multimodal streams, achieving high accuracy—some above 98~\%—and enabling real-time feedback with minimal latency \cite{4.1.f-8743159, 4.1.g-article, 4.1.e-article, 4.1.p-article, 4.1.q-sabbella2024evaluatinggesturerecognitionvirtual}.

The tracking method used in most cases relies on vision-based input, which ranges from basic webcam feeds \cite{4.1.f-8743159} to structured video datasets \cite{4.1.b-article} and  MediaPipe Holistic landmark detection \cite{4.1.a-unknown, 4.1.k-article}. The systems operating with hardware devices make use of the Oculus Quest 2 or Meta Quest 2  \cite{4.1.h-article, 4.1.i-article, 4.1.j-article} to record hand position, orientation, and grip parameters for immediate evaluation and sEMG signal acquisition to detect muscle-based gestures \cite{4.1.m-8623035}. The analysis of gesture detection systems also includes evaluations for hand skeletonization, posture recognition, and dual-hand interaction design \cite{4.1.p-article, 4.1.o-article, 4.1.l-article}. Most research relies on data-driven modeling, yet specific conceptual frameworks stress user-centered evaluation and gesture datasets standardization \cite{4.1.d-article, 4.1.q-sabbella2024evaluatinggesturerecognitionvirtual, 4.1.r-inbook}.

The majority of gesture recognition systems included real-time feedback mechanisms, which allowed users to correct their mistakes and enhance their learning outcomes \cite{4.1.p-article, 4.1.o-article, 4.1.e-article}. The applications show gesture recognition outcomes through on-screen visual cues or VR immersive interfaces, improving user interaction and engagement \cite{4.1.m-8623035, 4.1.q-sabbella2024evaluatinggesturerecognitionvirtual}.

Several Unity-based implementations applied gesture models to 3D spaces that included characters and animations together with gameplay elements \cite{4.1.p-article, 4.1.q-sabbella2024evaluatinggesturerecognitionvirtual}. These systems functioned as gamified VR teaching tools for children, using Oculus tracking and Unity rendering for SL instruction \cite{4.1.h-article, 4.1.i-article, 4.1.j-article}.

Most systems use visual feedback, but they don't include haptic feedback, like vibrations or physical sensations, to guide users. Even though some systems use gloves or VR controllers, these devices aren't used to give touch-based responses. Adding this kind of physical feedback could improve learning, especially for activities that rely on movement and touch \cite{4.1.h-article,4.1.i-article}.

Despite strong performance in controlled environments, most systems showed limitations related to hardware reliance, data collection variability, and sensitivity to environmental conditions \cite{4.1.h-article, 4.1.l-article, 4.1.m-8623035}. These factors indicate that there is still a requirement for scalable and adaptive models that perform well with users from different backgrounds in various environments and languages \cite{4.1.o-article, 4.1.q-sabbella2024evaluatinggesturerecognitionvirtual}.

\subsection{Interactive VR Environments for Practice}

Virtual reality environments with immersive features enable SL students to practice through role-playing, simulation, and gamified interaction \cite{A-alam2023aslchampvirtualreality, N-10025347}. Learners can enhance their sign accuracy by practicing gestures in meaningful contexts because these environments provide experiential cues and task completion feedback \cite{D-9072479}. Multiple VR-based systems for SL education use game mechanics and scenario progression to boost user engagement and motivation \cite{J-article, M-inproceedings}. The user's progress, task success, and virtual object interaction respond directly to gesture input in these environments, thus establishing a natural feedback system between learning and doing \cite{Q-article, Y-AdamoVillani20063dSL}.

VR applications in SL now integrate multiplayer features that enable users to engage in real-time interaction, cooperative activities, and dialogue-based practice \cite{berrezueta2025immersive}. The systems recognize SL as a social communication method, establishing virtual environments that mimic natural communication scenarios. Research involving DHH participants shows that multiplayer platforms enhance collaborative learning while enabling non-verbal communication and peer engagement through visual methods such as gestures and avatar interactions \cite{4.2.d-inproceedings}. However, in several cases, SL was used alongside or replaced by text chat or speech-to-text tools, especially when users did not share the same SL \cite{4.2.c-unknown}.

The broader educational VR environments have shown through multiplayer design that this approach leads to better immersion and collaboration and improved contextual learning results \cite{4.2.a-inproceedings}. The direct implementation of multiplayer SL interaction within VR learning systems would create more authentic communication experiences, which would help students develop fluent and responsive communication skills and social self-assurance \cite{4.2.b-inbook}.

\subsection{Immersive Learning Through Gamification}

Several serious games for SL education demonstrate that gameplay elements, such as animations, scores, or avatar reactions, enhance memory retention by linking gestures to specific outcomes \cite{J-article, A-alam2023aslchampvirtualreality}. This immediate feedback reinforces motor memory and comprehension, making learning more effective and engaging. Studies show that when signs trigger in-game actions or puzzles, learners perceive the experience as more meaningful and enduring than traditional methods \cite{N-10025347, Q-article}.

VR systems have been developed where American, British, and Iranian SLs are practiced through mini-games, object interaction, and goal-oriented puzzles \cite{J-article, A-alam2023aslchampvirtualreality, N-10025347, Q-article}. Some environments are designed to teach domain-specific vocabulary, such as mathematical signs, by embedding them in contextual fantasy scenarios \cite{Y-AdamoVillani20063dSL}. Others provide real-time feedback on gesture shape, timing, and execution, helping users internalize corrections through repeated gameplay \cite{C-inproceedings, M-inproceedings}. These implementations demonstrate how serious games can facilitate SL practice while maintaining learner engagement through meaningful task design, immediate reinforcement, and immersive interaction.

\subsection{Personalized Learning and Adaptive Systems}

Some approaches use environmental control and user-directed navigation to simulate personalized learning. For example, VR-based platforms for learning Brazilian Sign Language (LIBRAS) provided users with the ability to control camera angles, slow down animations, and replay signs at their own pace, which helped with visual self-pacing and independent practice without AI-driven adaptation \cite{4.4.b-7073231}. Similarly, systems incorporating learning style frameworks, like Kolb's experiential model, provide learners with multiple content formats (visual, auditory, reading) and let them choose non-linear learning paths based on preference \cite{4.4.c-9237895}. These approaches support learner autonomy but are based on manual customization rather than intelligent adjustment based on performance.

More advanced approaches propose the use of machine learning techniques for real-time adaptation. For instance, one AR/VR framework uses classification and clustering algorithms to track learner progress and recommend appropriate modules or feedback strategies on the fly. However, this system was not sign-language-specific \cite{4.4.d-ALANSI2023100532}. Nonetheless, there remains a gap between gesture-based feedback and truly adaptive, performance-driven learning pathways across the literature reviewed.

Current systems tend to focus more on autonomy rather than on intelligent adaptation. Few applications incorporate real-time analysis of gesture accuracy, fluency, or error trends to adjust lesson difficulty or suggest targeted practice. This gap indicates a clear opportunity for future work: integrating AI-driven user modeling and performance-based progression could significantly enhance the effectiveness of VR SL instruction by aligning content more closely with each learner's evolving needs.

\subsection{Accessibility and Inclusivity}

The visual aspects of VR environments match the communication requirements of DHH users by providing them with immersive learning experiences. Various systems show how gesture-based interaction combined with avatar-signed content enables more intuitive learning environments. A VR-based SL learning tool implemented motion-captured avatars alongside physical interaction through pinch gloves to let deaf children communicate naturally \cite{4.5.1-article}.

Several studies highlight user demand for accessible VR tools tailored to DHH learners. Survey-based research showed that 64~\% of deaf university students expressed interest in using VR for lectures, with 80~\% emphasizing the need for virtual SL interpreters. The same participants demonstrated better motivation and understanding when they engaged in virtual reality learning scenarios \cite{4.5.2-article}. 

Another study explored the development of multilingual sign translation in VR chatrooms, aiming to support more inclusive real-time communication. However, it remained at the prototype stage without empirical validation \cite{4.5.3-inbook}.

\section{Challenges}

\subsection{Hardware and Software Limitations}

VR-based SL learning systems experience limitations due to hardware and software restrictions. The literature shows an ongoing problem in finding scalable and affordable input and output devices that deliver high-fidelity performance \cite{4.1.o-article, 4.1.l-article}. Vision-based systems, while more accessible than sensor-laden systems, experience tracking problems and occlusion events along with latency issues and performance degradation when lighting conditions are poor \cite{4.1.q-sabbella2024evaluatinggesturerecognitionvirtual, 4.1.k-article}. Systems utilizing Leap Motion, webcams, or standalone VR headsets like Oculus Quest 2 report issues with limited tracking range, spatial resolution, and precision, particularly when recognizing detailed hand configurations or subtle finger movements \cite{4.1.i-article, 4.1.r-inbook}.

The accuracy-enhancing hardware solutions, including wearables, EMG sensors, and specialized gloves, bring their difficulties \cite{4.1.f-8743159}. Most gesture recognition methods demand complex hardware configurations while creating physical restrictions on users and generating variable signal quality, reducing system practicality, restricted scalability, minimal system interoperability, and restricted availability consistency \cite{4.1.m-8623035, 4.1.j-article}.

The software side faces additional restrictions because computational limitations make implementing advanced gesture recognition models in real-time systems difficult. The processing capabilities of standalone VR headsets prove insufficient to execute high-complexity models that use CNN-LSTM hybrids or transformer-based architectures \cite{A-alam2023aslchampvirtualreality, H6-inproceedings}.

Additionally, several studies highlight the lack of infrastructure and technical support in educational settings, which prevents the deployment of VR systems for widespread classroom use \cite{C-inproceedings, 4.1.d-article}. The operation and maintenance of VR tools demand specialized training for educators, which creates additional hurdles for adoption \cite{U-article}. Creating high-quality 3D signing avatars presents developmental challenges that require extensive resources and time while also complicating expanding support for multiple SLs and user groups \cite{M-inproceedings}.

Finally, despite the promise of web-based and AI-enhanced solutions, many VR applications still overlook fundamental accessibility needs, particularly for DHH users with intersecting impairments \cite{4.5.4-article}. Inclusive hardware-software design that addresses the needs of diverse learners remains largely underexplored.

\subsection{Accuracy of Gesture Recognition}

VR-based SL learning systems face accurate and consistent gesture recognition. Deep learning and motion tracking have produced significant progress, but many applications still face challenges when interpreting gestures in real-world immersive situations \cite{4.1.a-unknown, 4.1.b-article, H4-inproceedings}.

The fundamental problem stems from the challenging task of identifying static versus dynamic signs. Identifying dynamic gestures becomes challenging because of movement variations, hidden gestures, and hand motion overlap. The problem worsens because the SL community lacks sufficient standardized datasets that show regional and stylistic variations in signs as researched in Indian and Egyptian SL studies \cite{4.1.c-unknown, C-inproceedings}. The restricted nature of these systems to static vocabularies prevents them from supporting the complete range of continuous SL communication \cite{4.1.f-8743159, D-9072479}.

Vision-based systems show high sensitivity to lighting conditions, background elements, and hand position changes \cite{4.1.l-article, 4.1.n-article}. At the same time, commodity VR equipment from Oculus Quest and Leap Motion fails to detect precise finger movements and facial expressions \cite{4.1.o-article, 4.1.q-sabbella2024evaluatinggesturerecognitionvirtual, A-alam2023aslchampvirtualreality}.

Deep learning models, including CNNs,  LSTMs, and transformers, need extensive, well-annotated datasets for effective generalization. Current datasets show two major limitations: they either contain small amounts of data or do not provide enough time-based information for continuous SL recognition, resulting in overfitting and performance issues \cite{4.1.g-article, H5-inproceedings}. These constraints are especially limiting when systems depend on manual segmentation or weak supervision, as seen in several gesture pipelines \cite{4.1.p-article,D-9072479, H6-inproceedings}. Some studies have even found VR-based gesture recognition to underperform compared to traditional 2D methods \cite{J-article}.

Finally, the lack of integrated real-time feedback mechanisms remains a significant barrier to effective learning. Learners who do not receive prompt corrective information risk maintaining incorrect gestures, which create barriers to learning and skill development retention \cite{H-wen2024enhancingsignlanguageteaching, M-inproceedings}.

The solution demands three main components: dataset development with high quality and inclusivity, neural architecture hybridization for spatial-temporal feature detection, and real-time feedback integration \cite{B-10.1145/3411763.3451523, 4.1.h-article}.

\subsection{User Experience and Learning Effectiveness}

While VR can make learning more engaging, several studies show it doesn't constantly improve things like memory or long-term learning \cite{F-10536554}. Being immersed in a 3D, game-like environment may capture students' interest. Still, it can also be confusing and make learning harder when the instructions or tasks are too complex \cite{O-unknown, Q-article}.

Usability and personalization are critical yet often underdeveloped aspects of VR-based SL learning environments. Many systems fail to follow user-centered or iterative development processes, resulting in interfaces that confuse learners while contradicting their needs \cite{4.5.1-article}.

While interactive avatars and 3D elements can enhance visual clarity, their effectiveness hinges on high design standards, usability, and user acceptance \cite{4.4.b-7073231, M-inproceedings}. Moreover, standard VR platforms often lack accommodation for individual learning styles, affective preferences, and prior knowledge, essential for learner satisfaction and retention \cite{4.4.c-9237895, 4.4.d-ALANSI2023100532}. Although personalized VR/AR tools are conceptually promising, their implementation demands sophisticated analytics and adaptive systems that remain largely absent from current applications.

The implementation of gamification approaches in education produces varying levels of effectiveness. Research shows that motivational and engagement levels increase through gamification, yet superficial elements only provide meaningful learning benefits when psychological needs of autonomy, competence, and relatedness are specifically addressed \cite{J2-inproceedings, J3-article}. Contextual adaptation is critical; generic game elements may fail to support novice users or learners from diverse educational backgrounds \cite{O6-10.1007/978-3-319-09635-3_12, Q-article}.

The existing research demonstrates insufficient development of multiplayer and collaborative experiences that specifically address deaf and hard-of-hearing (DHH) users \cite{4.2.d-inproceedings}. The needs of DHH participants are not addressed through standard communication strategies, group dynamics, and co-located interaction designs, which restrict their involvement in social VR and educational simulations. In remote learning contexts, tools to support hearing parents in learning SL for their deaf children are especially scarce \cite{M-inproceedings}.

The effectiveness of VR-based SL learning relies on more than technological innovation, even though VR provides promising new interaction methods.

\subsection{Ethical and Cultural Considerations}

A new ethical concern is that AI systems may be biased, especially when personalizing learning. Research shows that people can have different experiences with AI based on gender, which risks reinforcing unfair treatment \cite{O4-10.1145/3640544.3645218}. Since SL learning relies on personalized feedback, biased AI could harm trust, exclude minority groups, and discourage users from continuing.

The ethical dimension of VR identity representation is a new, unexamined area that needs attention. Users tend to interact with avatars that do not match their cultural background and personal identity, which may result in feelings of exclusion and limited self-expression. Research indicates that identity satisfaction varies based on avatar customization options, suggesting that inclusive avatar design is essential for fostering belonging and user acceptance in educational VR \cite{4.1.j-article}.

The lack of qualified interpreters presents an infrastructural and ethical challenge to providing equal access to knowledge and academic participation. VR-based interpretation systems and avatar-led instruction can fill this gap, but only if designed with ethical inclusivity and linguistic accuracy in mind \cite{4.5.2-article}.

Undereducation and semi-literacy among deaf individuals stem from mainstream learning system exclusions and insufficient support for ongoing educational opportunities \cite{O2-article}. SL remains inaccessible as a standard school subject while fluent teacher availability remains limited, thus creating ongoing barriers to learning, especially in regions with insufficient resources \cite{K-article}.

Ethical design means more than supporting inclusion, representation, and equal access. Developers need to understand the diverse needs of deaf communities, including differences in language, gender, culture, and education. This calls for inclusive design, involving users in the development process, and checking AI systems for bias.
VR tools for SL learning will only reach their full potential if they are built with strong ethical standards. 

\section{Future Directions}

Several important areas still need more research and development to fully unlock the educational benefits of VR-based learning. 

\subsection{Advances in AI for Gesture Recognition}

The success of VR-based sign language learning depends largely on how accurately and quickly systems can recognize gestures. Future research should focus on using advanced deep learning models, such as CNNs, LSTMs, Transformers, and attention mechanisms, to better capture the movements and timing involved in continuous signing. There is also a strong need for large, diverse datasets that reflect real-world variation, including regional differences, facial expressions, and two-handed signs \cite{4.1.a-unknown, 4.1.c-unknown, 4.1.m-8623035}.

In addition, future systems should use multimodal input—combining hand motion with facial cues, voice, or muscle signals (EMG)—to improve recognition across different learning contexts \cite{4.1.n-article, A-alam2023aslchampvirtualreality}. Improving the speed and efficiency of these models on standard VR headsets is key for real-time feedback. Collaboration between AI experts and sign language linguists will be essential to ensure these systems reflect the full richness and diversity of sign languages \cite{B-10.1145/3411763.3451523, 4.1.g-article, 4.1.f-8743159, 4.1.p-article, H6-inproceedings}. 

\subsection{Integration with Augmented Reality (AR)}

Augmented Reality (AR) can enhance VR-based SL learning by enabling practice in real-world, context-rich environments. Using AR on mobile devices or smart glasses, learners can receive real-time feedback, visual overlays, and avatar demonstrations without needing a full VR setup. This makes learning more flexible and accessible, especially for mobile or blended learning approaches \cite{4.4.d-ALANSI2023100532, A-alam2023aslchampvirtualreality}.

For instance, learners could get immediate feedback while signing in everyday situations, such as ordering food or asking for directions, making applying skills in real life easier. AR could also help parents and teachers learn SL alongside deaf children in daily routines, promoting more inclusive and family-centered learning experiences \cite{D-9072479, H-wen2024enhancingsignlanguageteaching}

\subsection{Open-Source Development and Community Initiatives}

Many existing VR sign language systems were developed as standalone projects, limiting their reuse and scalability. To increase global impact, future work should focus on open-source platforms, modular toolkits, and standardized APIs for gesture recognition, avatars, and feedback. Publicly available datasets and reproducible models will support broader adoption, validation, and continuous improvement \cite{M-inproceedings, 4.5.1-article, 4.5.2-article}.

Involving Deaf and Hard-of-Hearing (DHH) users as co-creators ensures these tools meet real-world needs and reflect linguistic and cultural diversity. Future platforms should also allow user-generated content, allowing learners and educators to build and share custom lessons and learning paths \cite{4.5.4-article, A-alam2023aslchampvirtualreality}.

\subsection{Ethical AI and Representation in Sign Language VR}

Ethical design must be central to developing VR SL learning tools to ensure fairness and inclusivity. Future systems should address bias in AI models through fairness audits, diverse training data, and precise feedback mechanisms. Avatar design should also reflect various identities, including different body types, skin tones, genders, and signing styles \cite{4.1.j-article, 4.5.2-article, 4.5.4-article}.

Developers should follow participatory design practices involving DHH users, educators, and interpreters at every stage. Clear ethical guidelines for data use, privacy, and accessibility are essential when designing for children or underserved communities. Future research should also prioritize culturally responsive content and expand support for SL, particularly in low-resource settings \cite{O2-article, O4-10.1145/3640544.3645218, K-article}.

\section{Recommendations for Developers}

Based on the analysis of recent research and system implementations in VR-based SL learning, we provide the following recommendations to developers aiming to build inclusive, effective, and scalable learning environments:

\subsection{Prioritize Inclusive and Participatory Design}
\begin{itemize}
    \item Involve DHH individuals throughout the design and evaluation process to ensure cultural relevance, accessibility, and usability.
    \item Allow for extensive avatar customization to support identity expression and increase user comfort and representation across age, ethnicity, gender, and disability.
    \item Ensure compatibility with assistive technologies (e.g., screen readers, closed captions) to support multi-disability access.
    \item Audit AI models for bias based on gender, language, or signing style to avoid reinforcing inequities.
\end{itemize}

\subsection{Enhance Gesture Recognition with Robust, Multimodal AI}
\begin{itemize}
    \item Use hybrid deep learning models (e.g., CNN-LSTM, 3D CNNs with attention) capable of processing static and dynamic gestures.
    \item Train models on diverse datasets reflecting real-world variation in signing styles, regional dialects, and continuous sentence structures.
\end{itemize}

\subsection{Design Real-Time Feedback Mechanisms}
\begin{itemize}
    \item Implement immediate visual or auditory cues to help learners self-correct gestures.
    \item Use gamified reinforcement, such as animated reactions or scoring systems, to increase motivation.
    \item Integrate haptic feedback (e.g., vibration) to support kinesthetic reinforcement of signs.
\end{itemize}

\subsection{Support Personalization and Adaptive Learning}
\begin{itemize}
    \item Incorporate AI-based learner modeling to adapt content difficulty and exercises based on performance.
    \item Offer differentiated feedback strategies tailored to learning styles or goals (e.g., expressive fluency vs. receptive accuracy).
\end{itemize}

\subsection{Optimize for Cost-Effective and Scalable Hardware}
\begin{itemize}
    \item Design for low-cost, standalone headsets (e.g., Meta Quest 2) or mobile devices to reduce barriers to access.
    \item Ensure performance on mid-range processors by optimizing model size and latency.
    \item Minimize hardware dependency (e.g., gloves or EMG sensors) to ensure broader deployment.
\end{itemize}

\subsection{Ensure Open Access and Reusability}
\begin{itemize}
    \item Release gesture datasets, source code, and VR environments under open licenses to encourage collaboration and transparency.
    \item Document best practices and evaluation metrics to foster replicability and longitudinal study comparisons.
\end{itemize}

Implementing these recommendations will help developers close gaps in accessibility, learning effectiveness, and inclusivity and drive progress in VR-based sign language education.

\section{Limitations of This Review}
While this review aimed to provide a comprehensive and structured synthesis of the current VR-based sign language education landscape, several methodological limitations must be acknowledged.
First, publication bias may have influenced the results, as studies reporting successful or innovative outcomes are more likely to be published than those with inconclusive or negative findings. This could have skewed the overall interpretation of effectiveness and feasibility in favor of more optimistic representations.
Second, the language scope was limited to English-language publications, excluding potentially relevant studies published in other languages. Given the global nature of sign language communities and technological development, this introduces a geographic and linguistic bias that may have overlooked valuable contributions from non-English-speaking regions.
These limitations suggest that the findings of this review should be interpreted with caution and considered part of an evolving body of research that would benefit from continued meta-analytical work, multilingual inclusion, and participatory analysis frameworks.

\section{Conclusion}

This paper has comprehensively reviewed the current state, opportunities, and challenges of using Virtual Reality (VR) for sign language (SL) education. We identified five key application areas through systematic analysis—gesture recognition, interactive practice, gamified learning, personalized instruction, and accessibility—and assessed their strengths and limitations.

Our findings show that VR, especially when paired with AI-powered gesture recognition, offers promising solutions to key challenges in SL learning, such as the shortage of qualified instructors, low engagement, and limited feedback. Gamification and avatar-based interaction further support motivation and retention.
However, several barriers remain. Hardware costs, limited gesture recognition accuracy, and the lack of adaptive learning features restrict scalability. Ethical and cultural concerns—including algorithmic bias and underrepresentation of diverse SLs—must also be addressed to ensure true inclusivity.

Future systems must focus on robust datasets, participatory and user-centered design, and AI-driven personalization to move the field forward. Building open-source, affordable, and culturally responsive tools is key to expanding access, especially in underserved regions.
Realizing the full potential of VR in SL education will require close collaboration among technologists, educators, linguists, and Deaf and Hard-of-Hearing (DHH) communities. With inclusive and thoughtful design, VR can become a powerful tool to enhance language learning, promote accessibility, and support learners worldwide.

\section*{Acknowledgment}

This research was financially supported by the TUM Campus Heilbronn Incentive Fund 2024 of the Technical University of Munich, TUM Campus Heilbronn. We gratefully acknowledge their support, which provided the essential resources and opportunities to conduct this study. 

\balance
\bibliographystyle{elsarticle-num-names}
\bibliography{VR_SL}

\end{document}